\documentclass[9pt,twocolumn,twoside]{opticajnl}
\journal{opticajournal} 

\setboolean{shortarticle}{false}

\usepackage{lineno}
\usepackage{bm}

\title{Triad of equivalence theorems for the radiant intensity of partially coherent beams on scattering}

\author[1,*]{Yi Ding}
\author[2,3]{Daomu Zhao}
\affil[1]{School of Physical Science and Technology, Southwest Jiaotong University, Chengdu 610031, China}
\affil[2]{Zhejiang Key Laboratory of Micro-nano Quantum Chips and Quantum Control, School of Physics, Zhejiang University, HangZhou 310058, China}
\affil[*]{yding2021@swjtu.edu.cn}
\affil[3]{dmz123@zju.edu.cn}
\begin{abstract}
By using Laplace's method for double integrals and the so-called beam condition obeyed by a partially coherent beamlike light field, we report the equivalence theory (ET) of partially coherent beams on scattering for the first time. We present the necessary and sufficient condition for the two scattered fields that have the same normalized radiant intensity distribution when Gaussian Schell-model beams whose effective beam widths are much greater than the effective transverse spectral coherence lengths are scattered by Gaussian Schell-model media. We find that the condition contain three implications, and each of them corresponds to a statement of an ET of radiant intensity in a scattering scenario, which exposes the concept of a previously unreported triad of ETs for the radiant intensity of partially coherent beams on scattering. We also find that the existing ET of plane waves on scattering, which only asserts that two scatterers with scattering potentials' correlations whose low-frequency antidiagonal spatial Fourier components are identical, essentially is merely the first member of our triad of ETs, while the other two hidden important members are completely ignored. Our findings are crucial for the inverse scattering problem since they contribute to avoid possible reconstruction errors in realistic situations, where the light field used to illuminate an unknown object is a partially coherent beam rather than an idealized plane wave.
\end{abstract}

\begin{document}

\maketitle
In the potential scattering theory of electromagnetic radiation and light \cite{Wolf212, Wolf2, Olgawolf, Tong, TTWang}, a problem of practical importance for both scientific and engineering communities is the inverse scattering problem, i.e., reconstructing an unknown object from scattered data measured from the scattered field outside the object, which provides a contactless and nondestructive means to detect the internal structure of an unknown object. Since Wolf and Habashy in a classic paper published in 1993 pointed out that there are neither deterministic nor random nonscattering scatterers for all directions of incidence and proved the uniqueness of the inverse scattering problem within the validity of the first-order Born approximation \cite{WolfHa}, it has been a common assumption that the relationship between the structural characteristics of an unknown scatterer and the scattered radiation generated by it is strictly one-to-one matched in the solution of the inverse scattering problem \cite{FW, Gbur, ZKW, LWF, DCK, KVW, Li, Jia, DDD, TTWang}. However, such a assumption has been challenged as the ET of light waves on scattering is established \cite{Wangzhao}. The ET asserts that two scatterers with scattering potentials' correlation functions whose six-dimensional spatial Fourier transforms have the same low-frequency antidiagonal elements will scatter a light wave to produce identical spectral density distribution in the far zone. The physical reason for the equivalence is that the differences in the local values of the strength function of the scattering potential of the medium can be compensated by the differences in the local values of the normalized correlation coefficient of the scattering potential of the medium, and vice versa. The ET was then applied to prove the equivalence between random media and deterministic media \cite{dingwang}, and between continuous media and particulate media \cite{zhaowang}. Very recently, the ET has further been extended to the scattering scenario in which the scatterers are spatially anisotropic ones \cite{PAN, Tao}.

The ET is of great importance for the inverse scattering problem because it illustrates possible errors in the process of determining the internal structure of an unknown scatterer from the scattered-field data. To date, however, the understanding of the ET has been completely confined to an idealized situation in which the illumination is a spatially coherent plane wave, and no literature has reported the ET for the scattering of partially coherent beams, which are realistic optical fields in laboratory. One of the main reasons for the longstanding lack of an ET for partially coherent beams on scattering may be due to the complexity of the potential scattering itself of partially coherent beams, which makes it difficult to perform analytic calculations and thus to obtain the corresponding equivalent conditions even within the accuracy of the first-order Born approximation for the scattering of common Gaussian Schell-model beams by random media \cite{zhangyuanyuan, jianyang}, unless other additional approximation techniques are used. In this work, by using Laplace’s method for double integrals and the so-called beam condition obeyed by a partially coherent beamlike light field, we will derive the analytical expression for the scattered radiant intensity when a Gaussian Schell-model beam is scattered by a Gaussian Schell-model medium. Based on this, we will present the necessary and sufficient condition that the two scattered fields generated by Gaussian Schell-model beams whose effective beam widths are much greater than the effective transverse spectral coherence lengths on scattering from Gaussian Schell-model media may have identical normalized radiant intensity, and report a previously unexposed traid of ETs for the radiant intensity of partially coherent beams on scattering. We will illustrate our results by numerical examples.

\begin{figure}[htbp]
\centering
\includegraphics[width=7.8cm]{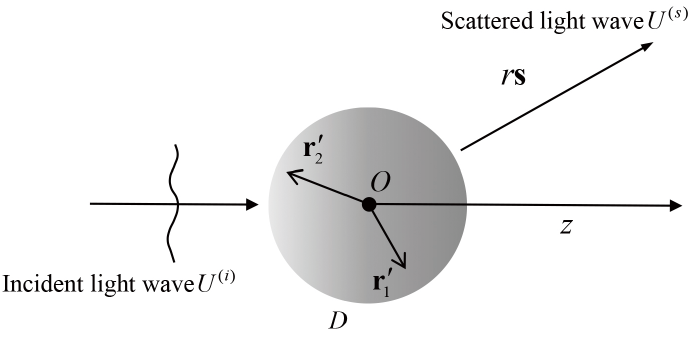}
\caption{Illustration of notations.}
\label{Fig 1}
\end{figure}

We assume that a monochromatic light field of frequency $\omega$, possessing a quite arbitrary structure, illuminates a linear scatterer, occupying a finite domain $D$ of space (see Fig. \ref{Fig 1}). The general way to represent the light field is to use an angular spectrum representation of plane waves propagating into the half-plane $z>0$ \cite{Olgawolf}
\begin{align}\label{infield}
    U^{(i)}(\mathbf{r},\omega)=\int_{|\mathbf{s}_{0\perp}|^2\leq1} a(\mathbf{s}_{0\perp},\omega)\exp{\bigl[ik\mathbf{s}_{0}\cdot \mathbf{r}\bigr]}d^{2}s_{0\perp},
\end{align}
where $k=\omega/c$ is the free-space wave number associated with frequency $\omega$, with $c$ being the speed of light in vacuum. $a(\mathbf{s}_{0\perp},\omega)$ is the amplitude of a plane wave propagating along a real unit vector $\mathbf{s}_{0}$, and $\mathbf{s}_{0\perp}$ is the two-dimensional vector obtained by projecting $\mathbf{s}_{0}$ onto the $x-y$ plane. $|\mathbf{s}_{0\perp}|\leq 1$ means that we neglect the contribution of evanescent plane waves for simplicity. 

In the potential scattering theory of light, the physical property of a deterministic scatterer is characterized by a potential function $F(\mathbf{r}^{\prime},\omega)$, which is related to the refractive index of the medium \cite{Wolf2}. Assume that the medium is a weak scatterer so that the scattering is analysed within the validity of the first-order Born approximation. The asymptotic approximation to the far scattered field as $kr\rightarrow\infty$ in a fixed direction $\mathbf{s}$ is given by \cite{Olgawolf}
\begin{align}\label{scafield}
    U^{(s)}(r\mathbf{s},\omega)=\frac{e^{ikr}}{r}\int_{|\mathbf{s}_{0\perp}|^2\leq1} a(\mathbf{s}_{0\perp},\omega)f(\mathbf{s},\mathbf{s}_{0},\omega)d^{2}s_{0\perp},
\end{align}
where
\begin{align}\label{scaamplitude}
f(\mathbf{s},\mathbf{s}_{0},\omega)=\int_{D}F(\mathbf{r}^{\prime},\omega)\exp{\bigl[-ik(\mathbf{s}-\mathbf{s}_{0})\cdot\mathbf{r}^{\prime}]}d^{3}r^{\prime}
\end{align}
is the the so-called scattering amplitude function.

We now consider the case where the incident field is not deterministic but a partially coherent light field and the scatterer also possesses random nature. The radiant intensity $J^{(s)}(\mathbf{s},\omega)$ of the scattered field, being the rate at which energy at frequency $\omega$ is radiated into the far zone, per unit solid around $\mathbf{s}$, can be calculated from \cite{Wolf2}
\begin{align}\label{radintensity}
    J^{(s)}(\mathbf{s},\omega)=r^{2}\langle U^{(s)*}(r\mathbf{s},\omega)U^{(s)}(r\mathbf{s},\omega)\rangle, \quad kr\rightarrow\infty
\end{align}
where $\langle\cdot\rangle$ denotes ensemble average and $*$ stands for complex conjugate. From now on we omit the dependence of all quantities on frequency $\omega$ for simplicity.

On substituting from Eq. (\ref{scafield}) into Eq. (\ref{radintensity}), the radiant intensity of the scattered field can be calculated as  
\begin{align}\label{radintensity1}
    J^{(s)}(\mathbf{s})&=\int_{|\mathbf{s}_{01\perp}|^2\leq1}\int_{|\mathbf{s}_{02\perp}|^2\leq1}A(\mathbf{s}_{01\perp},\mathbf{s}_{02\perp}) \notag \\ &\times  \widetilde{C}_{F}(-\mathbf{K}_{1},\mathbf{K}_{2})d^{2}s_{01\perp}d^{2}s_{02\perp},
\end{align}
where $A(\mathbf{s}_{01\perp},\mathbf{s}_{02\perp})=\langle a^{*}(\mathbf{s}_{01\perp})a(\mathbf{s}_{02\perp})\rangle$ is the angular correlation of the stochastic field, which can be calculated from the four-dimensional Fourier transform of the cross-spectral density function $W^{(0)}(\bm{\rho}_{1},\bm{\rho}_{2})$ of the stochastic field at a pair of points $\bm{\rho}_{1}=(x_{1},y_{1})$ and $\bm{\rho}_{2}=(x_{2},y_{2})$ in the source plane \cite{Wolf22} (the plane is set to pass through the center $O$ of the medium), with a form of
\begin{align}\label{angcorrefunction}
 A(\mathbf{s}_{01\perp},\mathbf{s}_{02\perp})&=\Bigr(\frac{k}{2\pi}\Bigl)^{4}\iint_{-\infty}^{\infty}W^{(0)}(\bm{\rho}_{1},\bm{\rho}_{2})\notag \\ &\times\exp\Bigl[{-ik(\mathbf{s}_{02\perp}\cdot\bm{\rho}_{2}-\mathbf{s}_{01\perp}\cdot\bm{\rho}}_{1})\Bigr]d^{2}\rho_{1}d^{2}\rho_{2},
\end{align}
and $\widetilde{C}_{F}(-\mathbf{K}_{1},\mathbf{K}_{2})=\langle f^{*}(\mathbf{s},\mathbf{s}_{01})f(\mathbf{s},\mathbf{s}_{02})\rangle$ is the correlation function of the scattering amplitude, which is computed from the six-dimensional spatial Fourier transform of the correlation function $C_{F}(\mathbf{r}_{1}^{\prime},\mathbf{r}_{2}^{\prime})=\langle F^{*}(\mathbf{r}_{1}^{\prime})F(\mathbf{r}_{2}^{\prime})\rangle$ of the scattering potential of the medium \cite{Wolf2}
\begin{align}\label{scaamplfunction}
   \widetilde{C}_{F}(-\mathbf{K}_{1},\mathbf{K}_{2})&=\iint_{D}C_{F}(\mathbf{r}_{1}^{\prime},\mathbf{r}_{2}^{\prime})\notag \\ &\times  \exp\Bigl[{-i(\mathbf{K}_{2}\cdot\mathbf{r}_{2}^{\prime}-\mathbf{K}_{1}\cdot\mathbf{r}_{1}^{\prime})}\Bigl]d^{3}r^{\prime}_{1}d^{3}r^{\prime}_{2},
\end{align}
where $\mathbf{K}_{1}=k(\mathbf{s}-\mathbf{s}_{01})$ and $\mathbf{K}_{2}=k(\mathbf{s}-\mathbf{s}_{02})$ are analogous to the momentum transfer vector of quantum mechanical theory of potential scattering \cite{Wolf2}.

In the following, for the stochastic light field, we consider a wide class of partially coherent beams, i.e., the well-known Gaussian Schell-model beam, which is an important extension of a laser beam operating in its lowest-order Hermite-Gaussian mode \cite{Wolf2}. The cross-spectral density function of a Gaussian Schell-model beam in the source plane is given by \cite{Wolf22}
\begin{align}\label{crospecfunction}
    W^{(0)}(\bm{\rho}_{1},\bm{\rho}_{2})=B_{0}\exp\Bigl[{-\frac{\bm{\rho}_{1}^{2}+\bm{\rho}_{2}^{2}}{4\sigma_{I}^{2}}}\Bigr]\exp\Bigl[{-\frac{(\bm{\rho}_{1}-\bm{\rho}_{2})^{2}}{2\sigma_{g}^{2}}}\Bigr],
\end{align}
where $B_{0}$ is positive constant. $\sigma_{I}$ and $\sigma_{g}$ represent the effective beam width and the effective transverse spectral coherence length of the beam, respectively.

For the scattering medium, without loss of generality, we will consider the so-called Gaussian Schell-model scatterer, which is commonly used to model many practical scatterers such as troposphere and confined plasmas \cite{Wolf2}. The correlation function of the scattering potential of a Gaussian Schell-model medium has the form \cite{Wolf2} 
\begin{align}\label{scapotenfunction}
    C_{F}(\mathbf{r}_{1}^{\prime},\mathbf{r}_{2}^{\prime})=A_{0}\exp\Bigl[{-\frac{\mathbf{r}_{1}^{\prime 2}+\mathbf{r}_{2}^{\prime 2}}{4\sigma_{S}^{2}}}\Bigr]\exp\Bigl[{-\frac{(\mathbf{r}_{1}^{\prime}-\mathbf{r}_{2}^{\prime})^{2}}{2\sigma_{\mu}^{2}}}\Bigr],
\end{align}
where $A_{0}$ is positive constant. $\sigma_{S}$ and $\sigma_{\mu}$ denotes the effective width of the strength of the scattering potential and the effective correlation width of the normalized correlation coefficient of the scattering potential of the medium, respectively. 

On substituting from Eqs. (\ref{crospecfunction}) and (\ref{scapotenfunction}) into Eqs. (\ref{angcorrefunction}) and (\ref{scaamplfunction}), respectively, and let us first change the variables of integration from $\bm{\rho}_{1}$ and $\bm{\rho}_{2}$ to $\bm{\rho}_{S}$ and $\bm{\rho}_{D}$ as well as from $\mathbf{r}^{\prime}_{1}$ and $\mathbf{r}^{\prime}_{2}$ to $\mathbf{R}_{S}$ to $\mathbf{R}_{D}$ by using the transformations
\begin{align}
  \mathbf{R}_{S}&=\frac{1}{2}(\mathbf{r}^{\prime}_{1}+\mathbf{r}^{\prime}_{2}), \quad \mathbf{R}_{D}=(\mathbf{r}^{\prime}_{2}-\mathbf{r}^{\prime}_{1}), \\
  \bm{\rho}_{S}&=\frac{1}{2}(\bm{\rho}_{1}+\bm{\rho}_{2}), \quad \bm{\rho}_{D}=\bm{\rho}_{2}-\bm{\rho}_{1},  
\end{align}
and we then perform the corresponding Fourier transforms and insert the results into Eq. (\ref{radintensity1}), and finally the radiant intensity of the scattered field can be written as 
\begin{align}\label{radintensity2}
    J^{(s)}(\mathbf{s})&=C_{0}\int_{|\mathbf{s}_{01\perp}|^2\leq1}\int_{|\mathbf{s}_{02\perp}|^2\leq1}\exp\Big[{-\frac{\sigma_{S}^{2}}{2}(\mathbf{K}_{1}-\mathbf{K}_{2})^{2}}\Bigl]\notag \\ &\times \exp\Big[{-\frac{\gamma^{2}}{8}(\mathbf{K}_{1}+\mathbf{K}_{2})^{2}-\frac{k^{2}\delta^{2}}{8}(\mathbf{s}_{01\perp}+\mathbf{s}_{02\perp})^{2}}\Bigl]\notag \\ &\times\exp\Big[{-\frac{k^{2}\sigma_{I}^{2}}{2}(\mathbf{s}_{01\perp}-\mathbf{s}_{02\perp})^{2}}\Bigl]d^{2}s_{01\perp}d^{2}s_{02\perp},
\end{align}
where
\begin{align}
    \frac{1}{\gamma^{2}}&=\frac{1}{4\sigma^{2}_{S}}+\frac{1}{\sigma^{2}_{\mu}} \label{constant1}, \\
    \frac{1}{\delta^{2}}&=\frac{1}{4\sigma^{2}_{I}}+\frac{1}{\sigma^{2}_{g}} \label{constant2}, \\
    C_{0}&=16\pi A_{0}B_{0}k^{4}\sigma_{I}^{2}\gamma^{2}\sigma_{S}^{3}\delta^{3}\label{constant3} .
\end{align}

We note that an analytic expression for radiant intensity from Eq. (\ref{radintensity2}) is generally unavailable because $\mathbf{s}_{01}$ and $\mathbf{s}_{02}$ are entangled in the exponential terms containing $\mathbf{K}_{1}$ and $\mathbf{K}_{2}$. However, the situation will become quite different if we confine our attention to the commonly occurring case in laboratory, i.e., the effective beam width $\sigma_{I}$ is much greater compared to the effective transverse spectral coherence length of the beam $\sigma_{g}$. In this case, inspired by the method adopted by Dijk \textit{et al} to address the four integrations similar to those in Eq. (\ref{radintensity2}) in their work on the scattering of a Gaussian Schell-model beam by an uniform deterministic sphere in the framework of Mie theory \cite{Dijk, Dijk1}, we take $k\sigma_{I}\rightarrow\infty$ in the double integral over $\mathbf{s}_{01\perp}$ and use Laplace's technique \cite{Wong, Lopez}, which says that the following integral holds
\begin{align}\label{Laplace}
   \iint_{\Omega}e^{-tf(x,y)}g(x,y)dxdy&\sim\frac{\pi g(x_{0},y_{0})e^{-tf(x_{0},y_{0})}}{t\sqrt{\text{Det}\Big[H f(x_{0},y_{0})\Bigr]}}, t\rightarrow\infty,  
\end{align}
where $f(x,y)$ and $g(x,y)$ are two smooth enough functions in a bounded or unbounded convex two-dimensional domain $\Omega$ and $t$ is a large positive parameter, and ($x_{0},y_{0}$) is the point where the minimum of $f(x,y)$ is attained in $\Omega$, and $H f(x_{0},y_{0})$ stands for the Hessian matrix of $f(x,y)$, computed at the point ($x_{0},y_{0}$)
\begin{equation}\label{Hessianmatrix}
Hf(x_{0},y_{0})=\left[ 
\begin{array}{cc}
  \partial^{2}_{x}f(x,y)& \partial_{x}\partial_{y}f(x,y)\\  
  \partial_{y}\partial_{x}f(x,y)& \partial^{2}_{y}f(x,y)
\end{array} 
\right]_{x=x_{0},y=y_{0}}.
\end{equation}

Here the large parameter $t$ and the functions $f$ and $g$ are selected as the following expressions
\begin{align}\label{function1}
    t&=(k\sigma_I)^{2}, \\
    f(\mathbf{s}_{01\perp},\mathbf{s}_{02\perp})&=\frac{1}{2}(\mathbf{s}_{01\perp}-\mathbf{s}_{02\perp})^{2}, \\
g(\mathbf{s}_{01\perp},\mathbf{s}_{02\perp},\mathbf{s})&=C_{0}\exp\Big[{-\frac{k^{2}\delta^{2}}{8}(\mathbf{s}_{01\perp}+\mathbf{s}_{02\perp})^{2}}\Bigl] \notag\\&\times\exp\Big[{-\frac{\sigma_{S}^{2}}{2}(\mathbf{K}_{1}-\mathbf{K}_{2})^{2}-\frac{\gamma^{2}}{8}(\mathbf{K}_{1}+\mathbf{K}_{2})^{2}}\Bigl].
\end{align}
It is easy to see that $f(\mathbf{s}_{01\perp},\mathbf{s}_{02\perp})$ as a function of $\mathbf{s}_{01\perp}$ takes its minimum when $\mathbf{s}_{01\perp}=\mathbf{s}_{02\perp}$, and the determinant of the Hessian matrix of $f(\mathbf{s}_{01\perp},\mathbf{s}_{02\perp})$ at the point $(s_{02x},s_{02y})$ can be easily shown to equal unity. 

Finally, we use the beam condition for the beamlike field whose beam axis along $z$ direction \cite{Wolf22} to further deal with the remaining double integral of $\mathbf{s}_{02\perp}$. The beam condition asserts that only those plane waves whose directions fall within a narrow solid angle around $z$-axis contribute significantly to the angular spectrum integral of the beam, which allows us to extend the low and upper limits in the double integral of $\mathbf{s}_{02\perp}$ to negative infinity and positive infinity, respectively, and we invoke the first two terms of the binomial expansion for $s_{02z}=(1-s_{02x}^{2}-s_{02y}^{2})^{1/2}$, i.e., $s_{02z}\approx1-(s_{02x}^{2}+s_{02y}^{2})/2$. Equation (\ref{radintensity2}) finally can be analytically computed as
\begin{align}\label{radintensity6}
   J^{(s)}(\mathbf{s})&=\frac{C_{0}\pi^{2}}{tM^{2}}\exp\Bigl[{\frac{k^{4}\gamma^{4}}{4M^{2}}}\mathbf{s}_{\perp}^{2}\Bigr]\exp\Bigl[-k^{2}\gamma^{2}(1-s_{z})\Bigr],
\end{align}
where
\begin{align}
    M=k\sqrt{(\frac{1}{2}\delta^{2}+\frac{1}{4}\gamma^{2}s_{z})}.
\end{align}

Equation (\ref{radintensity6}) implies that the necessary and sufficient condition under which any two radiant intensity distributions produced by Gaussian Schell-model beams on scattering from Gaussian Schell-model media will be identical can be formulated as 
\begin{align}
  \gamma_{1}&=\gamma_{2}, \label{simplecondition2} \\ \delta_{1}&=\delta_{2}  \label{simplecondition3},
\end{align}
or, more explicitly from Eqs. ({\ref{constant1}}) and ({\ref{constant2}})
\begin{subequations}\label{finalcondition}
\begin{align}
   \frac{1}{4\sigma_{S1}^{2}}+\frac{1}{\sigma_{\mu1}^{2}}&=\frac{1}{4\sigma_{\mu2}^{2}}+\frac{1}{\sigma_{S2}^{2}}, \label{finalcondition1}\\
   \frac{1}{4\sigma_{I1}^{2}}+\frac{1}{\sigma_{g1}^{2}}&=\frac{1}{4\sigma_{g2}^{2}}+\frac{1}{\sigma_{I2}^{2}} \label{finalcondition2},
\end{align}
\end{subequations}
where $\sigma_{I}\gg\sigma_{g}$ and $1/4\sigma_{I}^{2}+1/\sigma_{g}^{2}\ll k^{2}/2$. The latter results from the requirement of the beamlike field.

Equation (\ref{finalcondition}) is a primary result which gives the specific restrictions for the structural characteristics of both the scattering media and the illuminating beams to produce the same normalized radiant intensity distribution when Gaussian Schell-model beams are scattered by Gaussian Schell-model media. It requires not only two scatterers with scattering potentials' correlation functions whose six-dimensional spatial Fourier transforms have the same low-frequency antidiagonal elements, but also two illuminating beams with cross-spectral density functions whose four-dimensional spatial Fourier transforms have identical low-frequency antidiagonal elements. The physical reason for the same normalized radiant intensities generated by Gaussian Schell-model beams on scattering from Gaussian Schell-model media is that the trade-off between the contributions of the illuminating beam's spectral density and of its spectral degree of coherence is attained, and meanwhile the trade-off between the contributions of the medium's density function and normalized correlation coefficient of the scattering potential of the scatterer is also attained. Actually, Eq. (\ref{finalcondition}) contains three implications, and each of them corresponds to a statement of an ET of radiant intensity in a scattering picture, and we call them a triad of ETs for the radiant intensity of partially coherent beams on scattering. 

The first implication is the possibility that the scattered radiant intensities may have the same distribution throughout the far zone when a partially coherent beam is scattered from two random media with different correlation characteristics. Thus the first member of our ETs is outlined as follows: \textit{a Gaussian Schell-model beam on scattering from two entirely Gaussian Schell-model media will produce the scattered fields that have identical normalized radiant intensity distribution in the far zone provided that the characteristics of the scattering medium obey the second condition in Eq. (\ref{finalcondition1}}).

The second implication is the possibility that two partially coherent beams with different cross-spectral density functions can produce the same radiation intensity distribution in the far zone when scattered from a random medium. Thus the second member of our ETs is formulated as follows: \textit{two entirely different Gaussian Schell-model beams on scattering from a Gaussian Schell-model medium will produce the scattered fields that have identical normalized radiant intensity distribution provided that the characteristics of the illuminating beams are restricted by the first condition in Eq. (\ref{finalcondition2}}).

The third implication is the possibility that the distributions of the scattered radiant intensity are identical throughout the far zone when two partially coherent beams with different cross-spectral density functions are scattered from two entirely different random media. Thus the last member of our ETs is stated as follows: \textit{two different Gaussian Schell-model beams on scattering from two different Gaussian Schell-model media will produce the scattered fields that have identical normalized radiant intensity distribution provided that the characteristics of the scatterer and the incident beams are subject to their own restrictive conditions in Eqs. (\ref{finalcondition1}) and (\ref{finalcondition2})}.

Clearly, the connotations of the ET of partially coherent beams on scattering are much richer than those of plane waves on scattering. The ET of plane waves on scattering, which only says that two scatterers whose low-frequency antidiagonal spatial Fourier components of scattering potentials' correlation functions are identical, essentially is merely the first member of our triad of ETs, whereas the second and third members have never been exposed. What's more, from Refs. \cite{CW, WC}, we know that two beams possessing structural characteristics given in Eq. (\ref{finalcondition2}) and propagating in free space may produce the same radiant intensity in the far zone. The second and third members of our triad of ETs tell us that such two beams may still be able to produce identical radiant intensity in the far zone of the scatterers that scatter these two beams. This is somewhat unexpected because it implies that the perturbation of the light beam by the random scattering medium doesn't disturb the trade-off between the contributions of the light beam's spectral density and of its spectral degree of coherence.

\begin{figure}[ht]
\centering
\includegraphics[width=7.8cm]{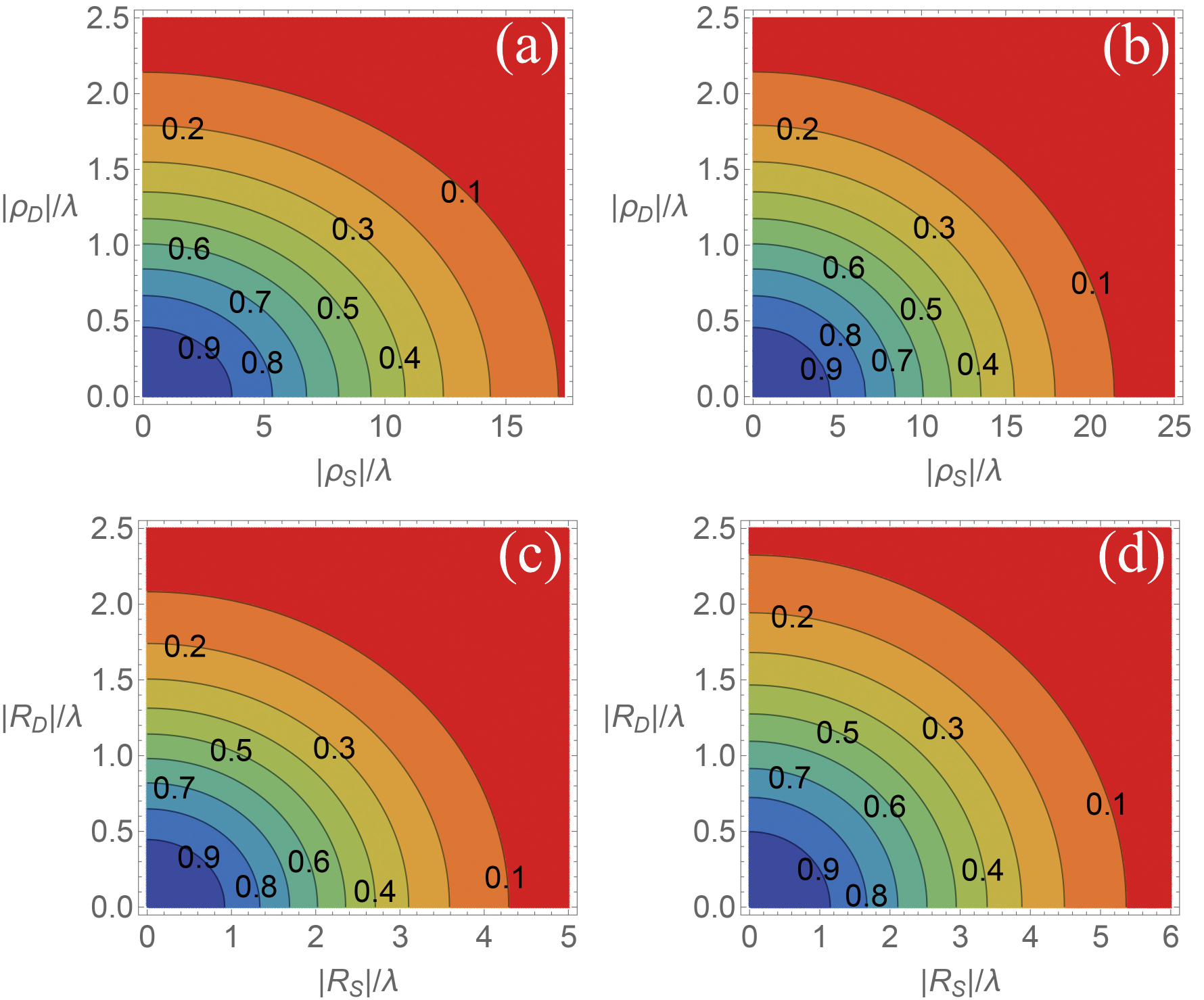}
\caption{Distributions of different correlation functions. The parameters for calculations are $\lambda=632.8 nm$, $k=2\pi/\lambda$, $A_{0}=B_{0}=1$. (a) $\sigma_{I}=8\lambda$, $\sigma_{g}=\lambda$; (b) $\sigma_{I}=8\sqrt{401}\lambda/\sqrt{257}$, $\sigma_{g}=0.8\sqrt{401}\lambda/\sqrt{257}$; (c) $\sigma_{S}=2\lambda$, $\sigma_{\mu}=\lambda$; (d) $\sigma_{S}=3\lambda$, $\sigma_{\mu}=12\sqrt{149}\lambda/149$.}
\label{Fig 2}
\end{figure}

In what follows, we will illustrate our results by some numerical examples. The cross-spectral density functions of two different illuminating beams are plotted in Figs. \ref{Fig 2}(a) and (b), while the correlation functions of the scattering potentials of two different media are shown in Figs. \ref{Fig 2}(c) and (d). The effective beam widths and the effective transversal spectral coherence lengths of the beams are not chosen arbitrarily but are based on the restrictions in Eq. (\ref{finalcondition2}). The same is true of the effective widths of the strength of the scattering potentials and the effective correlation widths of the normalized correlation coefficients of the scattering potentials of the media. Figure \ref{Fig 2} shows that the distributions of the cross-spectral density functions of the two beams are different and those of the correlation functions of the two media are also different, even though Eq. (\ref{finalcondition1}) and (\ref{finalcondition2}) hold.

\begin{figure}[t]
\centering
\includegraphics[width=7.8cm]{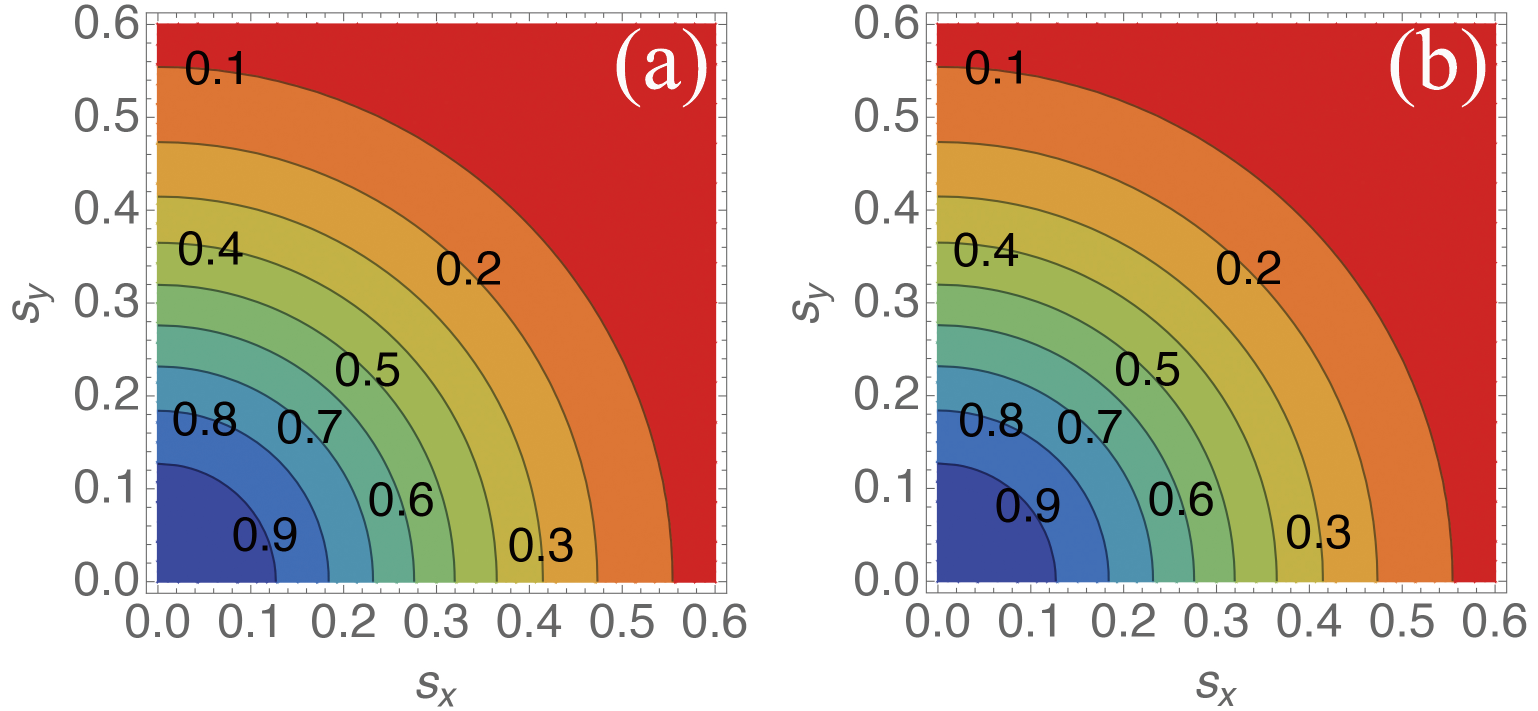}
\caption{Distributions of identical normalized radiant intensity. The parameters for calculations are (a) $\sigma_{I}=8\lambda$, $\sigma_{g}=\lambda$, $\sigma_{S}=2\lambda$, $\sigma_{\mu}=\lambda$; (b) $\sigma_{I}=8\sqrt{401}\lambda/\sqrt{257}$, $\sigma_{g}=0.8\sqrt{401}\lambda/\sqrt{257}$, $\sigma_{S}=3\lambda$, $\sigma_{\mu}=12\sqrt{149}\lambda/149$.}
\label{Fig 3}
\end{figure}

We are now in a position to focus on the radiant intensity generated by Gaussian Schell-model beams on scattering from random media. Figure \ref{Fig 3}(a) plots the normalized radiant intensity of a Gaussian Schell-model beam with the cross-spectral density function plotted in Fig. \ref{Fig 2}(a) on scattering from a Gaussian Schell-model medium with the correlation function displayed in Fig. \ref{Fig 2}(c), whereas the normalized radiant intensity of a different Gaussian Schell-model beam in Fig. \ref{Fig 2}(b) scattered by another Gaussian Schell-model medium in Fig. \ref{Fig 2}(d) is shown in Fig. \ref{Fig 3}(b). As we see from Figs. \ref{Fig 2} and \ref{Fig 3}, two normalized radiant intensity distributions generated by Gaussian Schell-model beams on scattering from Gaussian Schell-model media will be identical provided that the structural properties of the scattering medium and of the incident field meet Eqs. (\ref{finalcondition1}) and (\ref{finalcondition2}), respectively. It is not difficult to observe that the other two ETs have also been reflected in Figs. \ref{Fig 2} and \ref{Fig 3}. Thus the numerical results agree well with our triad of ETs.

In summary, we have reported the ETs of the radiant intensity of partially coherent beams on scattering for the first time. By exploiting Laplace's method for double integrals and the so-called beam condition obeyed by a partially coherent beamlike light field, we have derived the analytical conditions for the two scattered fields that have identical normalized radiant intensity distribution when Gaussian Schell-model beams are scattered from Gaussian Schell-model media. The resulting conditions state a previously unreported triad of ETs for the radiant intensity of partially coherent beams on scattering. The existing ET in the potential scattering theory essentially is merely the first member of our triad of ETs, while the second and third members were completely lost. These two members formulate that two Gaussian Schell-model beams satisfying (\ref{finalcondition2}) may not only produce the same radiant intensity in the far zone of free space, but also still be able to produce identical radiant intensity in the far zone of the scatterers that scatter these two beams, which implies that the perturbation of the illuminating beam by the random scatterer doesn't disrupt the trade-off between the contributions of the illuminating beam's spectral density and of its spectral degree of coherence. Although our triad of ETs is mainly for scattering of Gaussian Schell-model beams whose effective beam widths are much greater than the effective transverse spectral coherence lengths, it does resolve the longstanding lack of the ET of partially coherent beams on scattering and clarifies the relationship between the ET and the structural characteristics of both the illuminating beam and the scattering medium. The same radiant intensity measured from the scattered fields outside the scatterers may be not only from different media but also from the illuminating beams with different coherence characteristics. Such a clarification is beneficial to avoid possible reversion errors in the realistic inverse scattering problem such as remote sensing, object detection, and medical imaging and diagnostics, where the light field falling on an unknown scattering object is a partially coherent light.

\textbf{Acknowledgement.}
Financial support was provided by National Natural Science Foundation of China (12204385, 12474299).

\end{document}